\documentclass[english,10pt,aps,prd,a4paper,preprintnumbers,floatfix,nofootinbib,showpacs,superscriptaddress, notitlepage,twocolumn]{revtex4-1}

\pdfoutput=1

\usepackage{graphicx}
\usepackage{rotate}

\usepackage[utf8]{inputenc}

\usepackage{amsmath}
\usepackage{amssymb}
\usepackage{float}
\usepackage{comment}

\usepackage{soul}

\usepackage[usenames,dvipsnames]{color}
\usepackage[colorinlistoftodos]{todonotes}

\usepackage{amsmath}
\usepackage{amssymb}
\usepackage[colorlinks=true,citecolor=blue,urlcolor=blue, pdfborder={0 0 0}]{hyperref}
\usepackage[normalem]{ulem}
\newcommand {\ignore}[1]{}

\definecolor{darkred}{rgb}{0.6,0,0}

\usepackage[english]{babel}
\usepackage{blindtext}

\definecolor{mightnightblue}{RGB}{25,25,112}
\definecolor{brown}{rgb}{0.59, 0.29, 0.0}

\begin{document}
\bibliographystyle{unsrt}   

\title{Malaria detection from RBC images using shallow Convolutional Neural Networks}
\author{Subrata Sarkar}
\email{subrotosarkar32@gmail.com}

\affiliation{Government College of Engineering \& Ceramic Technology, Kolkata }

\author{Rati Sharma}
\email{rati@iiserb.ac.in}
\affiliation{Department of Chemistry, Indian Institute
of Science Education and Research (IISER), Bhopal - 462066, Madhya
Pradesh.}

\author{Kushal Shah}
\email{kushals@iiserb.ac.in}
\affiliation{Department of Electrical Engineering \& Computer Science, Indian Institute
of Science Education and Research (IISER), Bhopal - 462066, Madhya
Pradesh.}

\begin{abstract}

The advent of Deep Learning models like VGG-16 and Resnet-50 has considerably revolutionized the field of image classification, and by using these Convolutional Neural Networks (CNN) architectures, one can get a high classification accuracy on a wide variety of image datasets. However, these Deep Learning models have a very high computational complexity and so incur a high computational cost of running these algorithms as well as make it hard to interpret the results. In this paper, we present a shallow CNN architecture which gives the same classification accuracy as the VGG-16 and Resnet-50 models for thin blood smear RBC slide images for detection of malaria, while decreasing the computational run time by an order of magnitude. This can offer a significant advantage for commercial deployment of these algorithms, especially in poorer countries in Africa and some parts of the Indian subcontinent, where the menace of malaria is quite severe.

\end{abstract}
\maketitle


\section{Introduction}
\label{sec:intro}

Malaria is one of the six most prevalent infectious diseases of the world. According to the 2019 World Malaria Report \cite{Geneva:WorldHealthOrganization2019}, it inflicts more than 200 million people worldwide and causes more than 400,000 deaths every year. Given the risk that such an infection poses to human life, fast and accurate diagnosis is of great significance for early detection and management of the disease. Malaria is a mosquito borne disease caused by the parasite \textit{Plasmodium}, which invades the bloodstream of a human host through a mosquito bite. The parasite then develops and multiplies at the expense of the red blood cells (RBCs) which act as its host.

\begin{figure}
\begin{centering}
\includegraphics[scale=0.12]{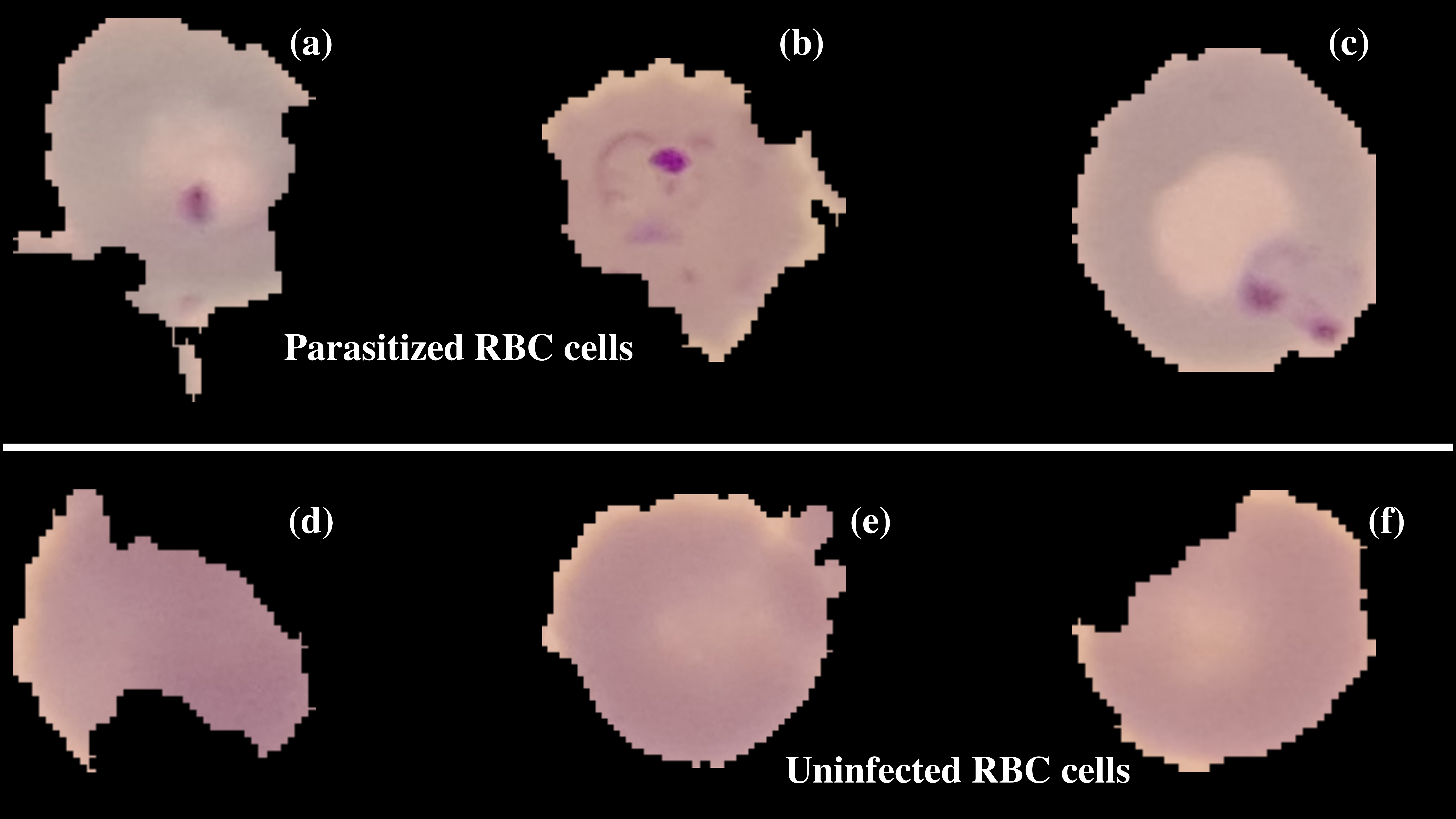}
\par\end{centering}
\caption{This figure shows the sample RBC images from our dataset \cite{Rajaraman}. (a), (b) and (c) are the parasitized RBC cells, whereas (d), (e) and (f) are the uninfected RBC cells. }
\label{Sample-Images}
\end{figure}

\begin{figure*}
\begin{centering}
\includegraphics[scale=0.6]{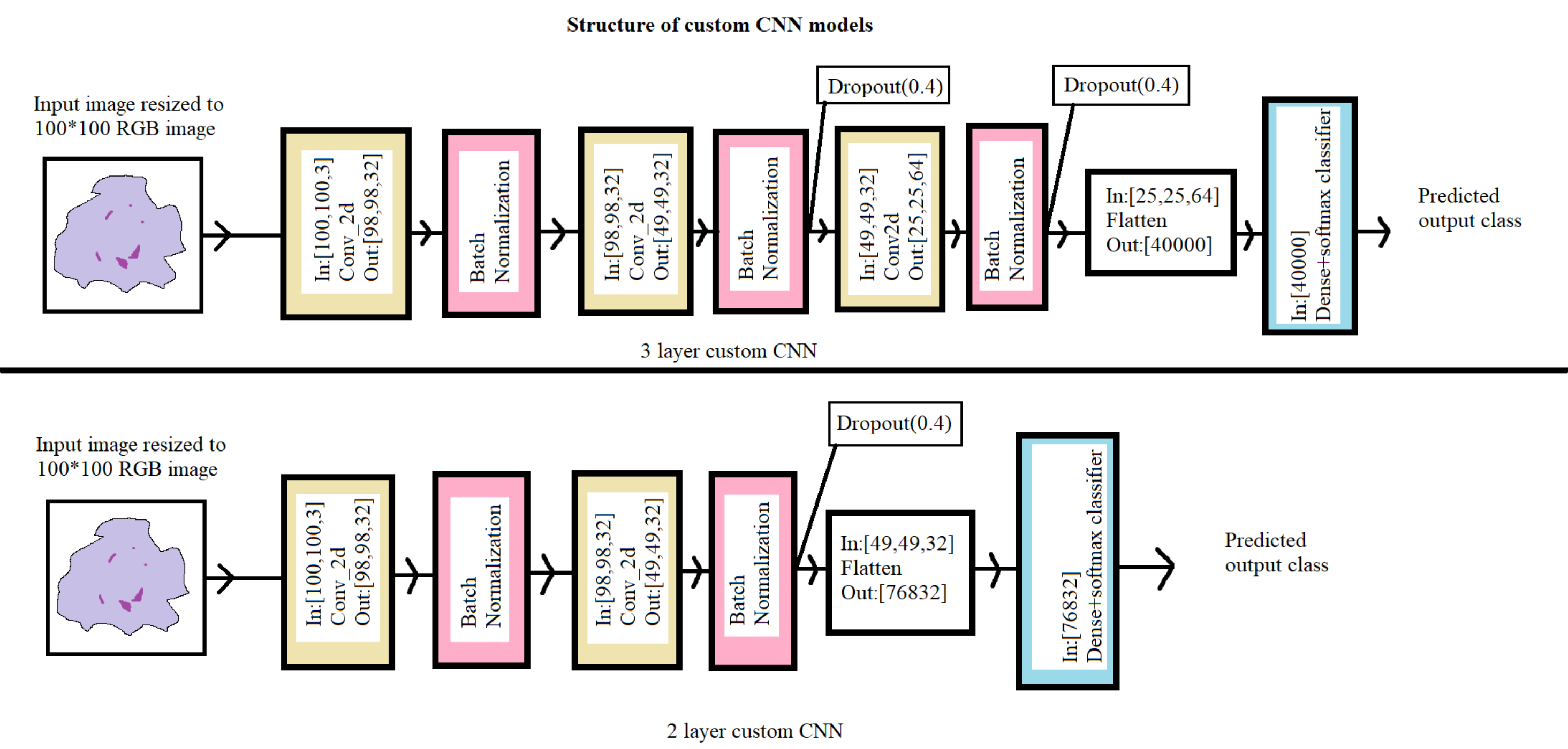}
\par\end{centering}
\caption{This figure shows the two shallow CNN models that we have used for classification of RBC images into infected/parasitized and uninfected categories. The top one is a 3-layer CNN model and the lower one is a 2-layer CNN model. As we show in this paper, both these shall CNN models are able to predict whether a given RBC image corresponds to a parasitized or uninfected cell with an accuracy as good as the conventional Resnet-50 and VGG-16 models.}
\label{CNN-arch}
\end{figure*}

The standard method of diagnosis of malarial infection is through the peripheral blood smear (PBS) test. This is an inexpensive technique wherein a patient's blood sample is smeared onto a glass slide creating a thin layer of RBCs. This is then fixed with methanol and treated with staining dyes (usually Giemsa or Leishman) and examined under a microscope for abnormalities within RBCs and for parasite quantification \cite{WHOWorldHealthOrganization2016, Brain2005}. The microscopic images of the RBCs can then be used to determine (i) the presence or absence of infection, (ii) the species of \textit{Plasmodium} that caused the infection and (iii) the stage of infection. 

However, manual examination of these images not only requires technical expertise in the field, but is also time consuming and prone to variability \cite{Constantino2015}. These limitations have led to the development of several different automated image processing and classification algorithms that distinguish and classify the morphology of cells within the peripheral blood smear images \cite{Diaz2009, Acevedo2019, Kratz2019, Merino2018, Rodellar2018}. For malaria detection, in particular, several different machine learning approaches have been proposed and developed which focus on specific aspects of classification (infected vs. non-infected cells, parasite species or stage of infection) \cite{Das2015, Poostchi2018, Rajaraman}. The input features on which the classification algorithms are applied are from one of three categories, namely, color, textural or morphologic \cite{Das2015a, Poostchi2018}. These features are then used to train the classification algorithms. Almost all the major machine learning based classification techniques have been applied to and tested on the PBS images, including, k-nearest neighbors (KNN) \cite{Prasad2012, Devi2018a, Diaz2007, Tek2010}, support vector machines (SVM) \cite{Diaz2009, Das2013, Linder2014, Molina2020}, naive Bayes \cite{Diaz2009, Das2013, Das2015a, Molina2020}, decision trees \cite{Molina2020, Khan2020} and deep learning methods \cite{Ross2006, Diaz2009, Razzak2015, Pamungkas2015, Devi2018, Gopakumar2018, Pan18, Acevedo2019, Dong2019, Delgado-Ortet2020}.   

In recent times, deep learning approaches have been the most successful in classification of medical images. These methods do not require manual feature selection or feature engineering. Complex convolutional neural network (CNN) architectures such as Resnet \cite{Resnet}, VGG-16 \cite{VGG-16} and Alexnet \cite{Krizhevsky2012} have several hyper parameters that can be tuned to produce extremely large and complex feature spaces in order to achieve the best accuracy. 


VGG-16, a CNN model proposed by Simonyan and Zisserman \cite{VGG-16}, achieves 92.7\% top-5 test accuracy in ImageNet, which is a dataset consisting of over 14 million images belonging to 1000 classes \cite{imagenet_cvpr09}. It was one of the best models submitted to the ImageNet Large Scale Visual Recognition  Challenge  (ILSVRC), 2014. This model made an improvement over AlexNet \cite{Krizhevsky2012} by replacing large kernel size filters (11 and 5 in the first and the second convolutional layer, respectively) with multiple $3\times 3$ kernel size filters one after another. The input to the VGG-16 is a $224\times224$ RGB image. which is then passed through a stack of convolutional layers having very small ($3\times3$) receptive filter fields. However, VGG16 suffers from two major drawbacks. Firstly, it is painfully slow to train. And secondly, the network architecture weights themselves are quite large, thereby increasing the load on the disk/bandwidth. 


Residual Networks or Resnet-50 \cite{Resnet} is a CNN model that was the winner of the ILSVRC 2015 challenge. Deep neural networks are generally hard to train due to the vanishing gradient problem. The Resnet CNN architecture tackled this problem by introducing the concept of skip connection, wherein, the original input is added to the output of a convolution block. This model has 5 stages, each with a convolution and an identity block. Each convolution block as well as each identity block has 3 convolution layers. 
Further, the Resnet-50 model has over 23 million trainable parameters. This improves the accuracy to a great extent, but also slows down training of the model compared to other comparatively shallower networks.

\begin{table*}
\begin{center}
\begin{tabular}{|c|c|c|c|c|c|c|c|c|c|c|}
\hline 
Models & Accuracy & Sensitivity & Specifivity & Precision & TP & TN & F1-score & MCC & Training time & Execution\tabularnewline
 &  &  &  &  &  &  &  &  &  & (single image)\tabularnewline
\hline 
\hline 
VGG-16 & 96.15\% & 94.82\% & 97.53\% & 97.54\% & 2652 & 2648 & 96.16\% & 0.9235 & 477.39s & 0.0145s\tabularnewline
\hline 
Resnet-50 & 94.07\% & 93.00\% & 95.10\% & 94.85\% & 2525 & 2660 & 93.92\% & 0.8815 & 2153.02s & 0.0117s\tabularnewline
\hline 
3 layered CNN & 95.32\% & 94.30\% & 96.34\% & 96.26\% & 2599 & 2655 & 95.27\% & 0.9066 & 701.05s & 0.0018s\tabularnewline
\hline 
3 layered CNN & 95.70\% & 93.21\% & 98.19\% & 98.09\% & 2569 & 2706 & 95.59\% & 0.9151 & 3219.65s & 0.002s\tabularnewline
(data augmentation) &  &  &  &  &  &  &  &  &  & \tabularnewline
\hline 
2 layered CNN & 90.82\% & 89.08\% & 92.56\% & 92.29\% & 2455 & 2551 & 90.66\% & 0.8169 & 615.38s & 0.002s\tabularnewline
\hline 
2 layered CNN & 95.61\% & 93.32\% & 97.90\% & 97.79\% & 2572 & 2698 & 95.51\% & 0.9131 & 3276.20s & 0.0019s\tabularnewline
(data augmentation) &  &  &  &  &  &  &  &  &  & \tabularnewline
\hline 
1 layered CNN & 49.96\% & 99.93\% & 00.00\% & 49.98\% & 2754 & 0 & 66.63\% & -0.0191 & 227.18s & 0.0029s\tabularnewline
\hline 
1 layered CNN & 69.38\% & 72.97\% & 65.78\% & 68.08\% & 2011 & 1813 & 70.44\% & 0.3885 & 3152.70s & 0.0026s\tabularnewline
(data augmentation) &  &  &  &  &  &  &  &  &  & \tabularnewline
\hline 
\end{tabular}
\par\end{center}
\caption{This table shows the details of our results obtained for various CNN models on the RBC image dataset. The accuracy and other metrics obtained for the shallow 3-layer CNN clearly are clearly as good as those obtained with the conventional Resnet-50 and VGG-16 models. The 2-layer CNN model also gives good performance when image data augmentation is used.}
\label{tab:Evaluation-metric-comparison}
\end{table*}

Although the models described above have been able to achieve high classification accuracy for malaria detection, they are like blackboxes even to machine learning experts as the complex feature spaces generated do improve accuracy but lose interpretability. This creates a lack of trust among the potential users (medical professionals) of such software, as they are unable to explain the reasons behind the classification of an image into a certain category. Our goal in this paper is to address this bottleneck, specifically for the problem of automated malaria detection, by developing a shallow CNN architecture that can give a classification accuracy that is as good as or better than what is achieved with Resnet or VGG models. A shallow CNN architecture also offers the advantage of better speed and cost effectiveness for commercial deployment of these algorithms, which can be a great advantage, especially in poorer countries in Africa and some remote parts of the Indian subcontinent, where the menace of malaria is quite severe.


\section{Methods}

To begin with, we tested the Resnet-50 and VGG-16 models using a 80:20 training-testing split on the publicly available malaria dataset \cite{Rajaraman}, and verified the earlier results (few sample RBC images shown in Fig. \ref{Sample-Images}). For implementing these algorithms, we used Keras with tensorflow backend in python. Next, we tried several different shallow CNN architectures on this dataset, and found two models that gave the best results. One is a shallow 3 layered network (Model 1) and the other is a 2 layered network (Model 2), as shown in Fig. \ref{CNN-arch}. We also report results of a 1 layered CNN model, for the sake of comparison.

In order to evaluate the efficacy of our models on the malaria dataset, we used various measures like accuracy, sensitivity, specificity, F1-score and MCC (Mathew's Corelation coefficient), listed below. Here, P stands for positive set (infected or parasitized), N stands for negative set (uninfected), TP stands for True Positive, TN stands for True Negative, FP stands for False Positive and FN stands for False Negative. 

\begin{equation}
\text{Accurary, }ACC=\frac{TP+TN}{P+N}    
\end{equation}

\begin{equation}
\text{Sensitivity/Recall, }TPR=\frac{TP}{TP+FN}    
\end{equation}

\begin{equation}
\text{Specifivity, }SPC=\frac{TN}{FP+TN}    
\end{equation}

\begin{equation}
\text{Precision, }PPV=\frac{TP}{TP+FP}    
\end{equation}


\begin{equation}
\text{F1 Score}=\frac{2TP}{2TP+FP+FN}    
\end{equation}

Matthews Correlation Coefficient : 
\begin{equation}
 MCC =  \frac{TP\cdot TN - FP\cdot FN}{\sqrt{(TP+FP)(TP+FN)(TN+FP)(TN+FN)}}
\end{equation}

\begin{figure}
\begin{centering}
\includegraphics[scale=0.3]{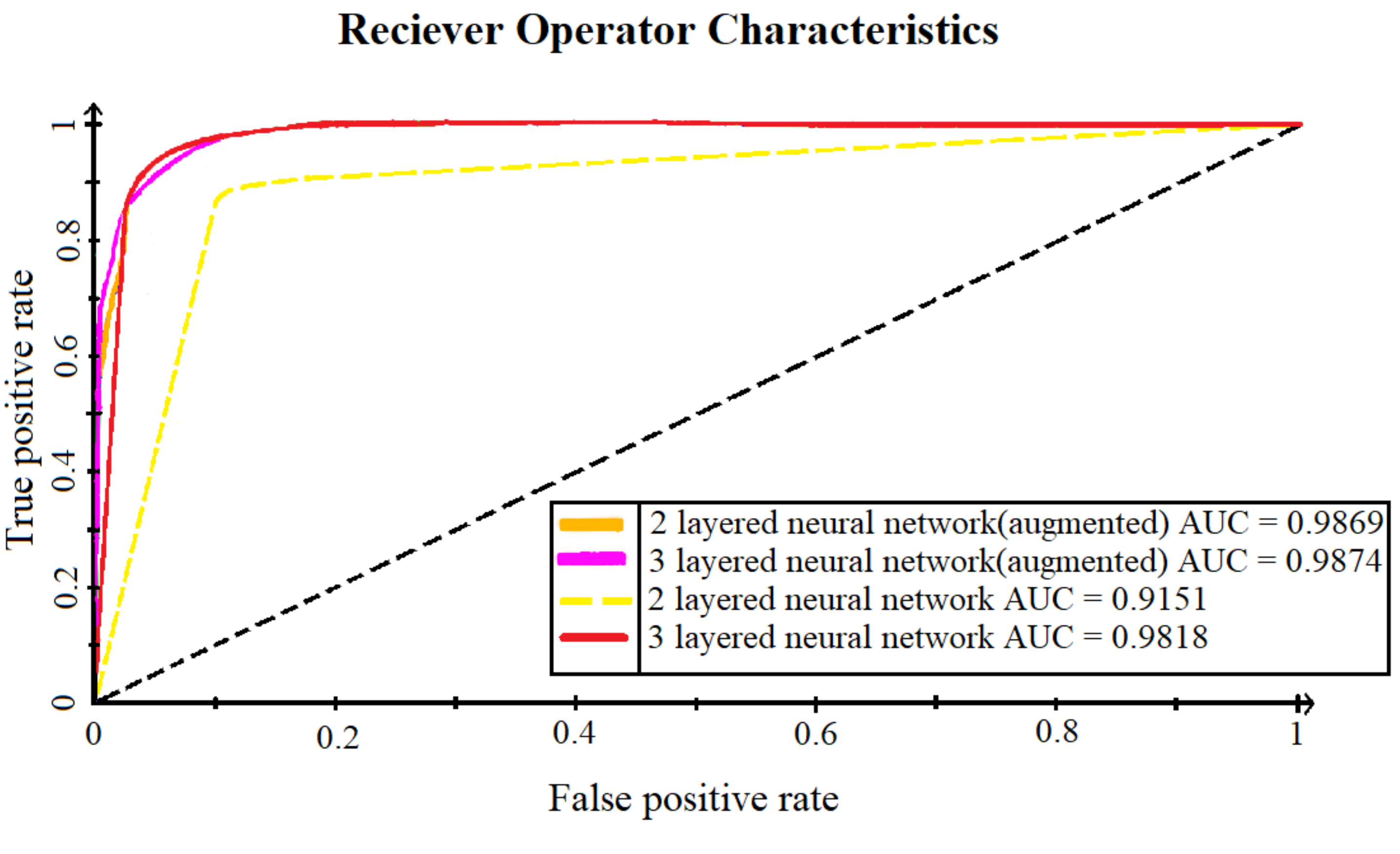}
\par\end{centering}
\caption{This figure shows the AUC-ROC [Area under the Curve for the Receiver Operating Characteristics] curve for our 3-layer and 2-layer CNN models with both the raw and augmented dataset. As can be clearly seen, the 2-layer CNN model does not give good results with raw dataset, but its performance becomes as good as that of the 3-layer CNN model with the augmented dataset.}
\label{AUC-ROC}
\end{figure}

\begin{figure*}
\begin{centering}
\includegraphics[scale=0.5]{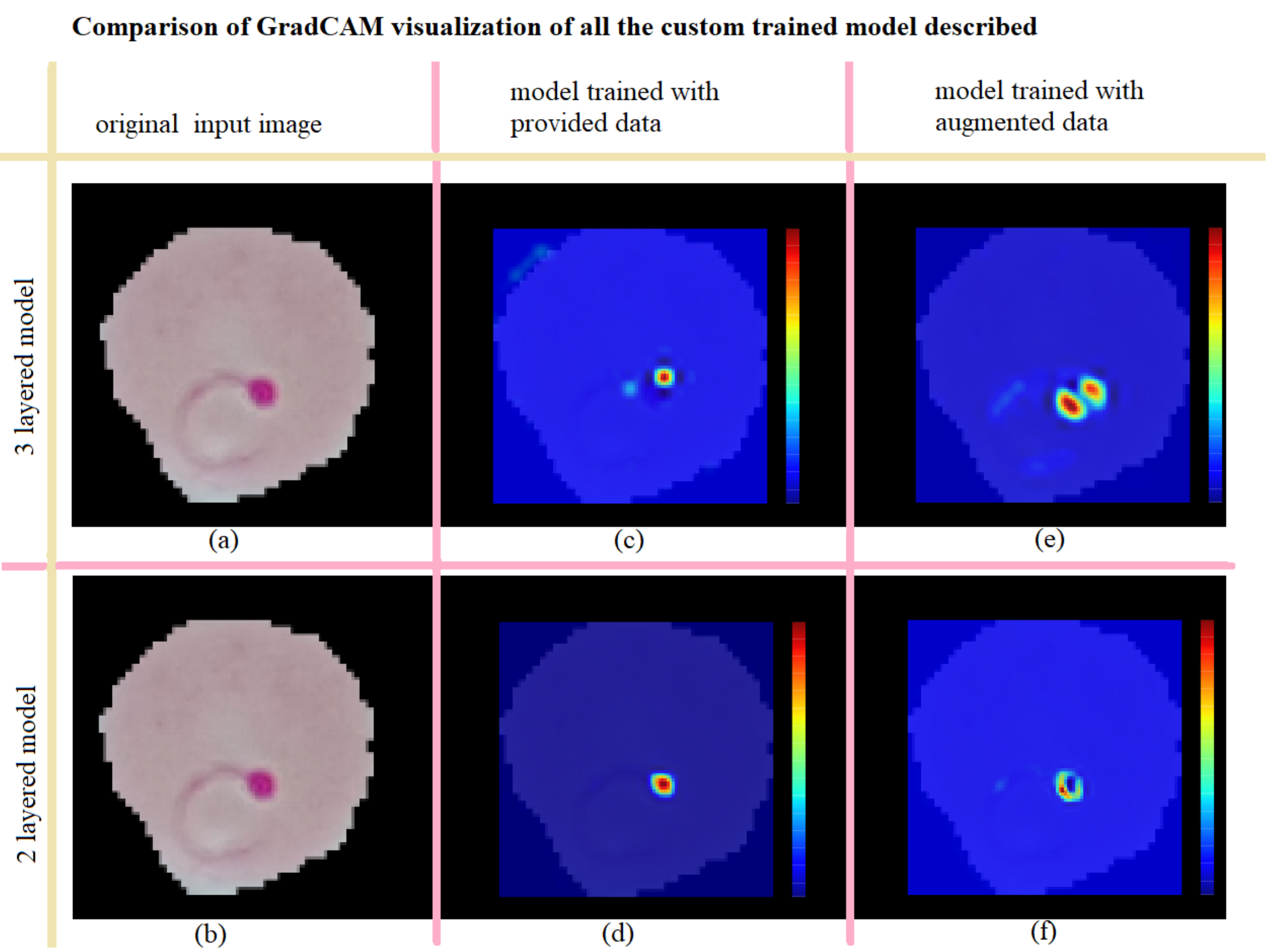}
\par\end{centering}
\caption{This figure shows the Grad-CAM images for one parasitized RBC cell for our shallow 3-layer and 2-layer CNN models for both the raw  and augmented dataset. As can be clearly seen, in all the four cases, the CNN model is able to correctly identify the location of the infection in the RBC cell image.}
\label{GradCam}
\end{figure*}

The seven metrics defined above can help us measure the model's performance from different perspectives. Precision or positive predictive value (PPV), for example, can be viewed as a measure of a classifier's exactness. A low precision also indicates a large number of False Positives. Sensitivity, also called recall or true positive rate (TPR), measures the proportion of positives that are correctly identified and it can be viewed as a measure of a classifier's completeness. A low recall indicates many False Negatives. Specificity (SPC), also known as true negative rate (TNR), measures the proportion of negatives that are correctly identified. F1-score encompasses both precision and recall and corresponds to the best accuracy when it reaches 1 and worst when it reaches 0. 
The measures described above can sometimes give a misleading picture if the dataset itself is unbalanced i.e., if there is a significant difference in the number of positive and negative samples. The Mathew's Correlation Function (MCC) is, however, free from these problems and provides an unbiased measure of a model's performance. MCC values can range from $-1$ to $+1$, where, a positive value closer to $+1$ indicates great performance. All of these measures have been determined for the various models trained on the RBC image dataset.

Another way of measuring the efficacy of any Machine Learning model is the Receiver Operating Characteristics (ROC) curve, which is a plot of the True Positive Rate (TPR) versus the False Positive Rate (FPR). The  Area Under the Curve (AUC) gives an estimate of the model's performance. In particular, the AUC of the ROC curve represents the degree of separability between the two classes (Positives and Negatives). A higher value of AUC means that the model is better at predicting the Negatives as Negatives and the Positives as Positives.

In the work described here, the 3 layered, the 2 layered and the 1 layered CNN models have all been trained for 60 epochs. In order to improve the testing accuracy, we have also used image data augmentation on the training set. Image data augmentation encompasses a wide range of techniques used to generate new training samples from the original ones by applying certain random modifications like rotation, scaling (zooming) and horizontal/vertical flips. The ImageDataGenerator class of Keras has been used to generate augmented images from the given RBC image dataset, which implements what is called an in-place/on-the-fly data augmentation.

Data Availability: 

The segmented cells from the thin blood smear slide images for the parasitized and uninfected classes are available at https://ceb.nlm.nih.gov/repositories/malaria-datasets/. The dataset contains a total of 27,558 cell images with equal instances of parasitized and uninfected cells \cite{Rajaraman}. 

\section{Results}

The main results of our work are presented in Table \ref{tab:Evaluation-metric-comparison}. As can be seen from this table, the 3 layered CNN model gives an accuracy of 95.32\%, which is almost similar to the accuracy obtained from the VGG-16 and Resnet-50 models. Further, the MCC score for the 3-layered model is 0.9066, which is slightly higher than these other two models. Usage of image augmentation slightly improves the efficacy of the 3-layered model in malaria image classification. This model also takes, on an average, only about 0.002 seconds to classify an image after being trained. 
This testing time is much lower than that required for the VGG-16 or the Resnet-50 models and therefore, can potentially provide a clear advantage during commercial deployment. Hence, the 3-layered CNN model clearly outperforms these other two models for the specific problem of malaria image classification.

Our next objective was to see if similar results can be obtained with a shallower, 2-layered CNN architecture. As can be seen from Table \ref{tab:Evaluation-metric-comparison}, this 2-layered architecture does not give very good results when the raw dataset is used. However, with the use of image augmentation on the training set, we obtain results that are as good as those obtained for the 3-layered CNN architecture, and with a similar testing time. We also note that the 3-layered and the 2-layered CNN architectures with image augmentation give MCC scores that are better than the Resnet-50 model and that are comparable to the VGG-16 model. We further reduced our model to a 1 layered CNN architecture, but as shown in the table, this generated very poor results. Hence, our primary finding in this work is that the shallow 2-layered CNN architecture (with image augmentation) is sufficient for the malaria image classification problem, and there is no need to use complex models like Resnet-50 or VGG-16. This conclusion is also corroborated by the AUC-ROC curve shown in Fig. \ref{AUC-ROC}.


Next, we wanted to check if our CNN architecture is doing the classification based on features that are relevant to the actual malarial infection or if some other aspect of the RBC image is being used during classification. We performed this analysis by applying the Gradient-weighted Class Activation Mapping (Grad-CAM) technique to the 3-layered and the 2-layered CNN models. Grad-CAM is a ``class-discriminative localization technique'' that highlights the region in the image that leads to a particular classification \cite{GradCAM}. Specifically, it achieves this by looking at the gradient flow into the last convolutional layer and then assigning weights to individual neurons for a particular classification. In order to verify the infection feature used for classification, as shown in Fig. \ref{GradCam}, we plotted the Grad-CAM images \cite{GradCAM} for a representative infected RBC for both the 3-layered and the 2-layered CNN models. Figures \ref{GradCam}a and \ref{GradCam}b show the raw image of the same infected RBC. Figures \ref{GradCam}c and \ref{GradCam}d show the Grad-CAM images for the model trained with the raw dataset and figures \ref{GradCam}e and \ref{GradCam}f show the Grad-CAM images obtained when the CNN model is trained with the augmented dataset. As can be clearly seen in all these four Grad-CAM images, both the 3-layered and the 2-layered CNN models are able to identify the actual location of the infection in the RBC image, which in the case of a malarial infection, is the center of the cell. We have also verified the same with the Grad-CAM images of other infected RBCs and found that this particular feature (infection location) is one of the primary aspects for classification.


\section{Conclusion and Discussion}

In summary, we have found two shallow CNN architectures which perform as well as VGG-16 and Resnet-50 architectures for thin blood smear RBC slide images for detection of malaria. Our proposed architecture has a significantly lower computational complexity, thereby lowering the computational run time by an order of magnitude. This enables these algorithm runs to require minimal computational infrastructure, which is especially relevant for commercial deployment in poorer parts of our world where malaria is quite prevalent. 

We believe that our results are relevant not just for detection of malaria but for other diseases as well of this nature, i.e. those which are detected using thin blood smear images of different kinds. We are currently unable to test our CNN architecture for other diseases due to lack of publicly available datasets, but hope these will become available in the future.

Apart from computational cost, a shallow CNN architecture also paves the way to explainable AI, since large networks have too many parameters, rendering the CNN to be a blackbox. This is especially relevant in the medical field, since medical diagnosis has a direct impact on human well-being and can also lead to ethical and legal issues. One of the reasons why Machine Learning algorithms have not seen widespread deployment for diagnosis purposes is the lack of explainability in Deep Neural Networks. Any medical doctor would naturally feel hesitant to use an algorithm to predict the presence of a disease in a person, if the algorithm cannot give the reason for its prediction (positive/negative). We hope that the development of shallow CNN architectures can address this issue and help in reducing the heavy burden on our medical infrastructure.  

One limitation of our work is that all the images in the dataset on which we have tested our CNN model were taken in the same way, under the same standardized microscopic conditions. But, there could be minor variations in the predictions when we compare images taken using different microscopic settings owing to differences in lighting, objective magnification, etc. However, we believe that our usage of data augmentation would have made our trained CNN models robust to these issues. In the future, we plan to acquire more malaria infected RBC image datasets from various laboratories in India, and check the accuracy of our CNN models, thereby validating them. It may happen that the current shallow models show a decrease in accuracy when images obtained from a wide variety of sources are used. In this case, we hope that methods of Knowledge Distillation \cite{Chen2018, Ho2020} will help us in reducing the complexity of our CNN models.

\begin{acknowledgments}
The authors would like to thank Ganapathy Krishnamurthi and Ishaan Gupta for very interesting discussions.
This work was financially supported through a grant from the Department
of Biotechnology, Government of India (File No. BT/PR33151/AI/133/21/2019). 
\end{acknowledgments}

\bibliographystyle{unsrt}
\bibliography{refs}

\end{document}